\documentclass[conference]{IEEEtran}
\IEEEoverridecommandlockouts
\usepackage{cite}
\usepackage{amsmath,amssymb,amsfonts}
\usepackage[ruled,vlined,linesnumbered]{algorithm2e}
\usepackage{graphicx}
\usepackage{textcomp}
\usepackage{xcolor}
\usepackage{pdflscape}
\usepackage{booktabs}
\usepackage{lipsum}
\usepackage{colortbl}
\usepackage{subcaption}

\usepackage[compact]{titlesec}
\usepackage{hyperref}
\usepackage{wasysym}
\usepackage{fancyhdr}
\hypersetup{
    colorlinks=true,
    linkcolor=blue,
    filecolor=blue,      
    urlcolor=blue,
    citecolor=blue
}

\def\BibTeX{{\rm B\kern-.05em{\sc i\kern-.025em b}\kern-.08em
    T\kern-.1667em\lower.7ex\hbox{E}\kern-.125emX}}

\begin{document}

\title{A Study of Password Security Factors among Bangladeshi Government Websites}

\author{\IEEEauthorblockN{Adil Ahmed Chowdhury, Farida Chowdhury, Md. Sadek Ferdous}
\IEEEauthorblockA{Department of Computer Science \& Engineering,
Shahjalal University of Science \& Technology, Sylhet, Bangladesh\\}
\IEEEauthorblockA{Email: adil.aj95@gmail.com, farida-cse@sust.edu, sadek-cse@sust.edu}}

\maketitle
\thispagestyle{fancy}
\pagestyle{fancy}
\fancyfoot[C]{}

\begin{abstract}
The Government of Bangladesh is aggressively transforming its public service landscape by transforming public services into online services via a number of websites. The motivation is that this would be a catalyst for a transformative change in every aspect of citizen life. Some web services must be protected from any unauthorised usages and passwords remain the most widely used credential mechanism for this purpose. However, if passwords are not adopted properly, they can be a cause for security breach. That is why it is important to study different aspects of password security on different websites. In this paper, we present a study of password security among $36$ different Bangladeshi government websites against six carefully-chosen password security heuristics. This study is the first of its kind in this domain and offers interesting insights. For example, many websites have not adopted proper security measures with respect to security. There is no password construction guideline adopted by many websites, thus creating a barrier for users to select a strong password. Some of them allow supposedly weak passwords and still do not utilise a secure HTTPS channel to transmit information over the Internet. 
\end{abstract}


\section{Introduction}
With almost a ubiquitous presence of the Internet, more and more services are provided online using web. A plethora of different web services have had a positive impact on almost every aspect of our human lives. Online services exhibit a number of additional advantages in comparison to the traditional mode of service delivery such as a global 24/7 access, cost savings, ease of use and flexibility, thus narrowing down the gap of accessibility hindrance \cite{online2020}.

Governments from all over the world are also offering an increasing amount of governmental public services via online, the so-called \textit{e-Government} initiative. With the \textit{Digital Bangladesh} initiative, the Government of Bangladesh has pledged to leverage ICTs (Information \& Communication Technologies) to facilitate a transformative change in every aspect of the life of a Bangladeshi citizen \cite{digitalBD2020}. Towards this goal, the Government of Bangladesh is offering an increasing number of online public service.

In recent times, different Bangladesh institutions and organisations are facing continuous cyber threats. The Bangladesh Bank Cyber Heist still remains the largest example, in terms of financial loss, of an online security breach for any organisation in the world \cite{BB2020}. Therefore, to counteract these security threats, we must guarantee the security of the offered public services. Indeed, many of these services handle sensitive information and hence, must ensure a restrictive access so that only authenticated and authorised users can access the correct personalised service. To enable this, every user must register at first and then login using an identifier (e.g. username, email address and so on) and a credential. There are a different forms of credentials utilised in online services, however, passwords remain the most widely used one. A recent study estimates that the number of passwords utilised for online services will be around $300$ billion by $2020$ \cite{passusage2020}. Unfortunately, even with its ubiquitous utilisation, password remains one of the major sources of security breaches for online services \cite{passstat2020}. That is why it is essential to study and understand different aspects of password security on different Bangladeshi government websites. Surprisingly, there is a pressing gap in this domain. In this paper, we report a study that aims to fill in this gap.

We have conducted a study analysing $36$ Bangladeshi government websites and their offered services against six chosen heuristics: password construction guidelines, password recovery mechanism, utilisation of CAPTCHA, security questions, utilisation of HTTPS and password strength meter. Using these six heuristics we aim to gauge the security status and preparedness of Bangladeshi government agencies and suggest a set of recommendations for any weakness found.

\vspace{2mm}
\noindent \textbf{Structure:} The rest of the paper is structured as follows. Section \ref{sec:related} describes related work whereas Section \ref{sec:pass} provides a brief background on password authentication along with a discussion of the chosen heuristics for the study. In Section \ref{sec:method}, we explain our methodology. Section \ref{sec:result} presents the results of this study along with their implications and outlines a set of recommendations. Finally, we conclude in Section \ref{sec:conclusion} with a hint of future work.

\section{Related Work}
\label{sec:related}

In this section, we explore a few related researches involving different aspects of passwords and other Bangladesh-focused security studies. Password research has been a widely studied research domain for around three decades or so with a number of influential researches. For example, a review of different types of passwords, their security mechanisms, possible attack methods and weaknesses as well as the password reuse issue can be found in \cite{klein1990foiling, raza2012survey, hoonakker2009password,das2014tangled}. There have been a number of researches exploring the influence of culture between different password aspects, such as on the memorability, utilisation and management, and the citizens of different countries \cite{yang2013analysis, constantinides2018cultural, aljahdali2013affect}. 

Despite passwords being a widely studied research domain in the field of security, the number of research papers in this domain targeting Bangladeshi citizens and web services are rare. Khan et. al presented a study on Bangladeshi people’s knowledge and security awareness about different security practices on the Internet \cite{khan2018security}. They took a survey of $1682$ Bangladeshi online users and asked them about different security practices like how often they changed their passwords, whether they wrote down or shared their passwords with others and so on. The result was then compared with the users of developed countries and unsurprisingly, they found security knowledge gaps among Bangladeshi users. In a similar study, authors in \cite{haque2013password} presented a comparative study of password construction and management strategies between the users of Bangladesh and developed countries. This study also reported that the most of the Bangladeshi users did not follow the general practices to make a password secure. Moniruzzaman et. al examined a number of vulnerabilities in different Bangladeshi websites \cite{moniruzzaman2019measuring} and found out that many Bangladeshi websites, mostly Bangladeshi government websites, were vulnerable against a number of attack vectors. 

These research works explored website securities, password awareness and security practices. However, to the best of the authors knowledge, there is no study exploring the utilisation of different password-related security factors among the government websites of Bangladesh. Such study would be extremely useful to gauge the security status and preparedness of Bangladeshi government agencies. In this paper, we aim to fill in this gap.

\section{Password authentication}
\thispagestyle{empty}
\label{sec:pass}
To provide personalised online services, the identity of every user must be verified using an authentication mechanism, a process of determining someone's identity \cite{goodrich2011introduction}. This is an essential mechanism in many real life scenarios when someone has to show an \textit{ID card} to prove their identities to access a service, e.g. borrowing a book from a library. However, authentication becomes indispensable in online services as there is no physical way to prove someone's identity. There are different authentications mechanisms which can be grouped in three different categories \cite{goodrich2011introduction}: something a user has (like a card or a key), something a user knows (like a secret phrase) and something the user is (bio-metrics such as a fingerprint). 

A password is an authentication mechanism that falls under the ``the something a user knows'' category. It is essentially a shared secret between the user and a service provider relying on the premise that this secret is not shared/disclosed to any other party and this is by which a user can be determined to access personalised services from a web service. There are a number of widely used authentication mechanisms for online services and mobile devices using passwords, bio-metrics such as fingerprints \cite{roddy1997fingerprint}, voice \cite{zhang2016voicelive} and retina \cite{marino2006personal}. However, password still remains the mostly used and deployed authentication mechanism as other techniques are relatively expensive to implement and maintain.

The security of a password is of paramount importance. If a password is not secure, a user may face a number of security breaches such as the respective account being hacked, data theft and so on \cite{ives2004domino}. To make a password secure, researchers have devised a number of techniques and guidelines. It has also been found that many users often write down their passwords in papers as they forget them frequently \cite{morris1979password}. However, writing down a password is a bad habit as it may compromise the security of the system even if a strong secure password is used. Sharing passwords with friends and families is another bad habit found among general users \cite{shay2010encountering} which a user should avoid doing. Users engage in such activities because of their of lack of knowledge on the importance of password security. Another important factor is how a password is transmitted from the web browser of a user to the service provider during the registration and login phases. 
Based on the above discussions, we have identified six crucial heuristics, presented below, which are important to assess different aspects of password security for any online service, including Bangladeshi government websites.
\begin{itemize}
\item \textbf{H1:} Password Construction Guidelines
\item \textbf{H2:} Password Recovery
\item \textbf{H3:} CAPTCHA
\item \textbf{H4:} Security Question
\item \textbf{H5:} HTTPS Channel
\item \textbf{H6:} Password Strength Meter
\end{itemize}

Next, we present a brief discussion of these heuristics.

\subsection{H1: Password Construction Guidelines}
There are a number of ways by which a secure and strong password can be constructed. These mechanisms are often outlined in a series of of requirements and recommendations known as \textit{password guidelines} \cite{furnell2007assessment}. Depending on the security requirements, different organisations utilise different password guidelines. For example, a password guideline could dictate the length of the chosen password or how different characters such as a small letter, capital letter, digits, or a special character can be mixed together to build a security password. The security of a password is often measured using \textit{entropy} which implies the difficulty of breaking a password using different attack methods such as guessing, brute-force, dictionary attacks and so on \cite{ma2010password}. Next, we explore different password constructions guidelines. 

A password with at least a length of $8$ characters with no composition (construction) requirement, is called \textit{Basic8 composition} \cite{komanduri2011passwords} and its entropy value is $18$ bits. A password with length at least $16$ characters with no composition requirement, is called \textit{Basic16 composition} \cite{komanduri2011passwords} which has an entropy of $30$ bits. Another composition requirement titled \textit{Dictionary8} requires a password to of at least length 8 with no dictionary words allowed. Its entropy value is $24$ bits. A composition technique with entropy $30$ bits where the length requirement is at least 8 does not allow users to use dictionary words and is constructed using upper-case, lower-case, digits, and special characters. This composition is called \textit{Comprehensive8} which has a similar entropy of \textit{Basic16}. A password with a higher entropy is considered more secure.

Displaying a password construction guideline on the website of the service providers during the registration phase (where a user chooses a password) can aid the users to choose a secure password and its enforcement ensures every password meets the minimum security requirements. Such a guideline also indicates the seriousness of the service provider with respect to the security. 

\subsection{H2: Password Recovery}
Since users need to maintain a number of passwords for many different websites, it is not surprising that they often forget their passwords. In fact, forgetting password is a common issue. To remedy this situation, service providers must provide a password recovery option. There are a number of ways this recovery mechanism can be initiated, e.g. by sending a reset link to the registered email or phone number or via another secure channel. 

\subsection{H3: CAPTCHA}
A CAPTCHA (Completely Automated Public Turing test to tell Computers and Humans Apart) is a security measure which is used to distinguish between a human and a computer program \cite{von2003captcha}. A CAPTCHA is designed in such a way that a person can pass the CAPTCHA test which requires to take decisions in a contoured or fuzzy environments. However, a computer program or script fails the test. CAPTCHA is integrated in the registration as well as password recovery phase of a web service to ensure that only valid users are registered to access their services, so as to guard against automated bots utilised by attackers \cite{rusu2010leveraging}. A CAPTCHA can be of different types such as text, image, audio, video and puzzle, each with their security strengths \cite{singh2014survey}.

\subsection{H4: Security Question}
To secure the recovery process, one of the widely used mechanisms is the option of one or more security questions and a user must answer the question(s) correctly before the password recovery option can be initiated. Such a question/answer mechanism can be used to create another layer of security to identify authentic users. Security questions and their answers are recorded during the registration phase and stored in the backend database. Later, when the user wants to recover the password, the chosen security question is asked again to match the answer, thereby cross-checking the authenticity of the user.

\subsection{H5: HTTPS Channel}
The service provider also has a crucial role to play for securing a password and other sensitive information while they are transmitted using a web protocol. HTTP (Hypertext Transfer Protocol) is a major web protocol used to transfer data, such as website contents or API calls, over the Internet between a web server and a client such as a browser, a mobile app or another web service. Unfortunately, HTTP is an insecure protocol \cite{rescorla2000http} meaning that every information is transmitted over an HTTP channel in plain-texts. This enables an attacker to observe the network traffic and get hold of sensitive information (such as passwords) transmitted over HTTP. To address this problem, a secure web protocol called \textit{HTTPS} (Hypertext Transfer Protocol Secure) has been devised. HTTPS is a secure extension of HTTP creating an encrypted channel using TLS/SSL on top of HTTP, thereby ensuring the confidentiality, integrity and authenticity of the web server as well as the contents transmitted between the web server and the client. Every web service should utilise the HTTPS channel to ensure that passwords and other sensitive information are never transmitted over an insecure (HTTP) channel.

\subsection{Password Strength Meter}
\begin{figure}[htbp]
\centering
\fbox{\includegraphics[trim=40mm 35mm 40mm 35mm,clip,width=0.40\textwidth]{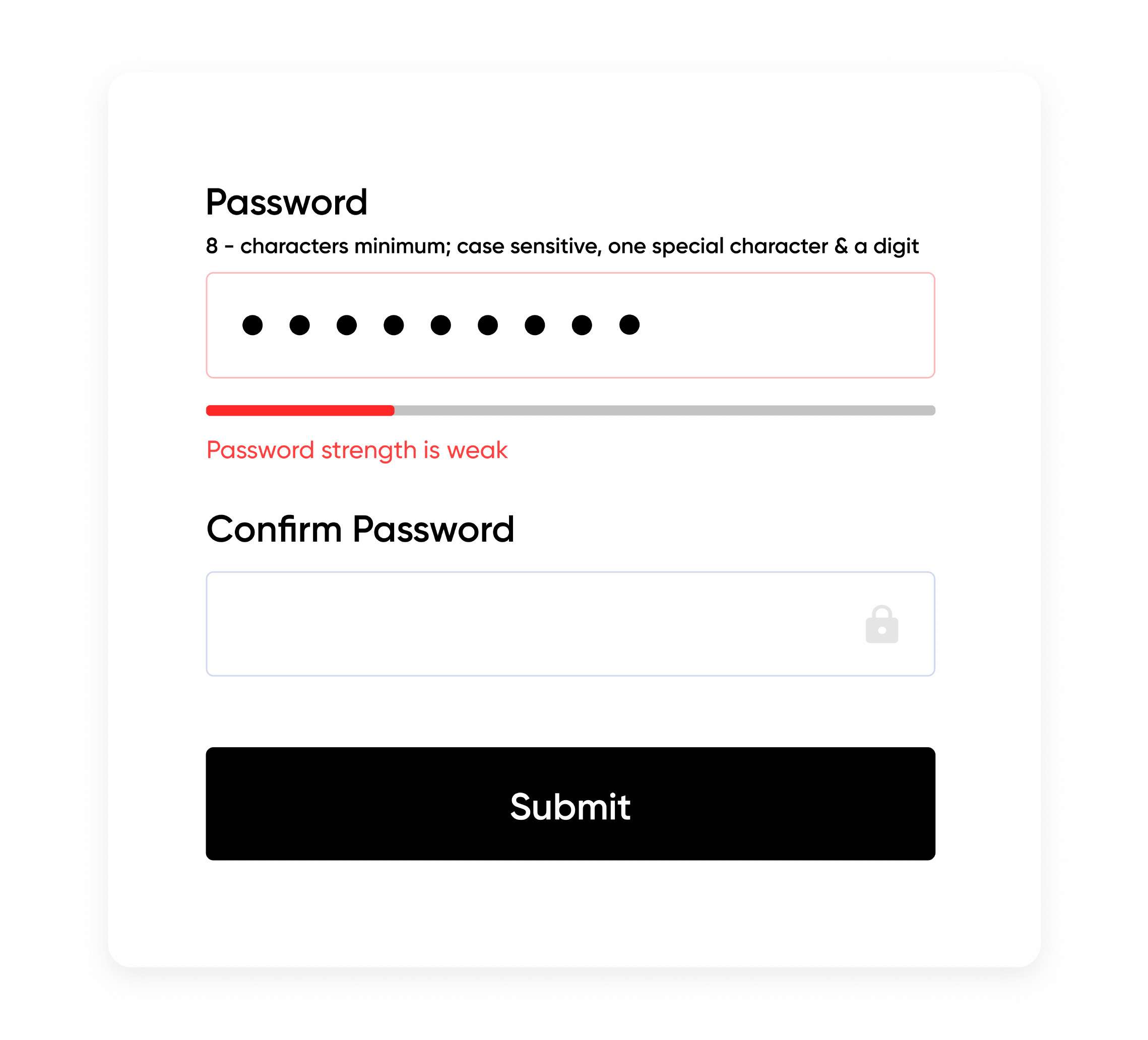}}
\caption{Password Strength Visualisation}
\label{fig:meter}
\end{figure}

One of the main reasons behind password hacking is the use of a weak password. Because of the lack of knowledge on security, a user may not be aware that (s)he is choosing a weak password. If a visual feedback could be provided when a password is created (during the registration phase), it would help the user to construct a strong password. A password strength meter is a visual representation of the strength of newly created password \cite{ur2012does} providing a visual cue, to the user, in real-time. This cue is provided using a combination of text and image where a password strength is categorised by \textit{weak}, \textit{medium} and \textit{strong} texts and a supplementary image reflects this category in different colors. Figure \ref{fig:meter} shows a sample of a password strength meter where a red color indicates that the given password is weak whereas green would refer it to be a strong password. It has been argued that a password strength meter can a great aid for the users \cite{de2014very} and that is why it has been included as one of our heuristics.

\section{Methodology}
\thispagestyle{empty}
\label{sec:method}
The methodology of our study is summarised in Algorithm \ref{algo:method}. At first, we have compiled a list of $120$ Bangladeshi Government websites (denoted with $\mathit{govList}$ in Algorithm \ref{algo:method}) by exploring the National portal of Bangladesh \cite{bnp2020} (an  integrated  online  portal recording all websites/services by the Bangladeshi government).

\setlength{\textfloatsep}{2pt}
\begin{algorithm}
\SetAlgoLined
\caption{Methodology} \label{algo:method}
\textbf{Input:} $\mathit{govList} \rightarrow$ all Govt. websites \\
\textbf{Output:} $\mathit{selList} \rightarrow$ selected websites with observations\\
\SetKwBlock{Begin}{}{}
\Begin(\textbf{Start})
{
  $\mathit{selList} = loginFunc(\mathit{govList})$\;
  \While{$x \in \mathit{selList}$}{
        \For{$i\gets1$ \KwTo $6$}{
            \uIf{$x$ satisfies $H_i$}{
                $\mathit{selList}.{H_i} = \mbox{``Yes''}$\;
            }
            \Else{
                $\mathit{selList}.{H_i} = \mbox{``No''}$\;
            }
            
        }
    }
  
  \SetKwFunction{login}{\textbf{loginFunc}}
  \SetKwProg{Fn}{function}{}{}
  \Fn{\login{$\mathit{govList}$}}{
    \While{$x \in \mathit{govList}$}{
        \uIf{$x$ has login functionality}{
            $\mathit{selList} = \mathit{selList} \cup x$\;
        }
  
    }
    \KwRet $\mathit{selList}$\;  
  }
}
\end{algorithm}

In the next step, we have shortlisted (denoted with $\mathit{selList}$ in Algorithm \ref{algo:method}) the $\mathit{govList}$ by visiting each website individually one by one and checking if they have any registration and login features so that a user can create an account for getting services from the website. The motivation behind this step is that not all websites will require registration and login features and hence, they do not need to deal with passwords and can be excluded. This process is explained in line $4$ and lines $14$ to $18$ in Algorithm \ref{algo:method}. At the end of this step, we have managed to find $36$ websites that meet our criteria. Among these $36$ websites, the majority ($30$, around $83\%$) provide services whereas five ($13.8$\%) provide information and one is a training website. The detailed information regarding these websites along with their URLs and purposes is presented in Table \ref{tab:findings}. 

In the final step, we have visited these $36$ websites one by one, created an account and investigated if each website satisfies the selected six heuristics as outlined in lines $5$ to $12$ in Algorithm \ref{algo:method}. Consequently, we have marked the heuristic for that respective website with a ‘Yes’ (line $8$ in Algorithm \ref{algo:method}). Otherwise the heuristic has been tagged with a 'No' (line $10$ in Algorithm \ref{algo:method}) with some additional comments.

\section{Results \& Discussion}
\thispagestyle{empty}
\label{sec:result}
In this section, we present and analyse the findings of our study (Section \ref{sec:result:subsec:analysis}) along with a discussion and a set of recommendations (Section \ref{sec:result:subsec:reco}).
\subsection{Analysis}
\label{sec:result:subsec:analysis}
\begin{figure}[h]
\centerline{\includegraphics[width=0.3\textwidth]{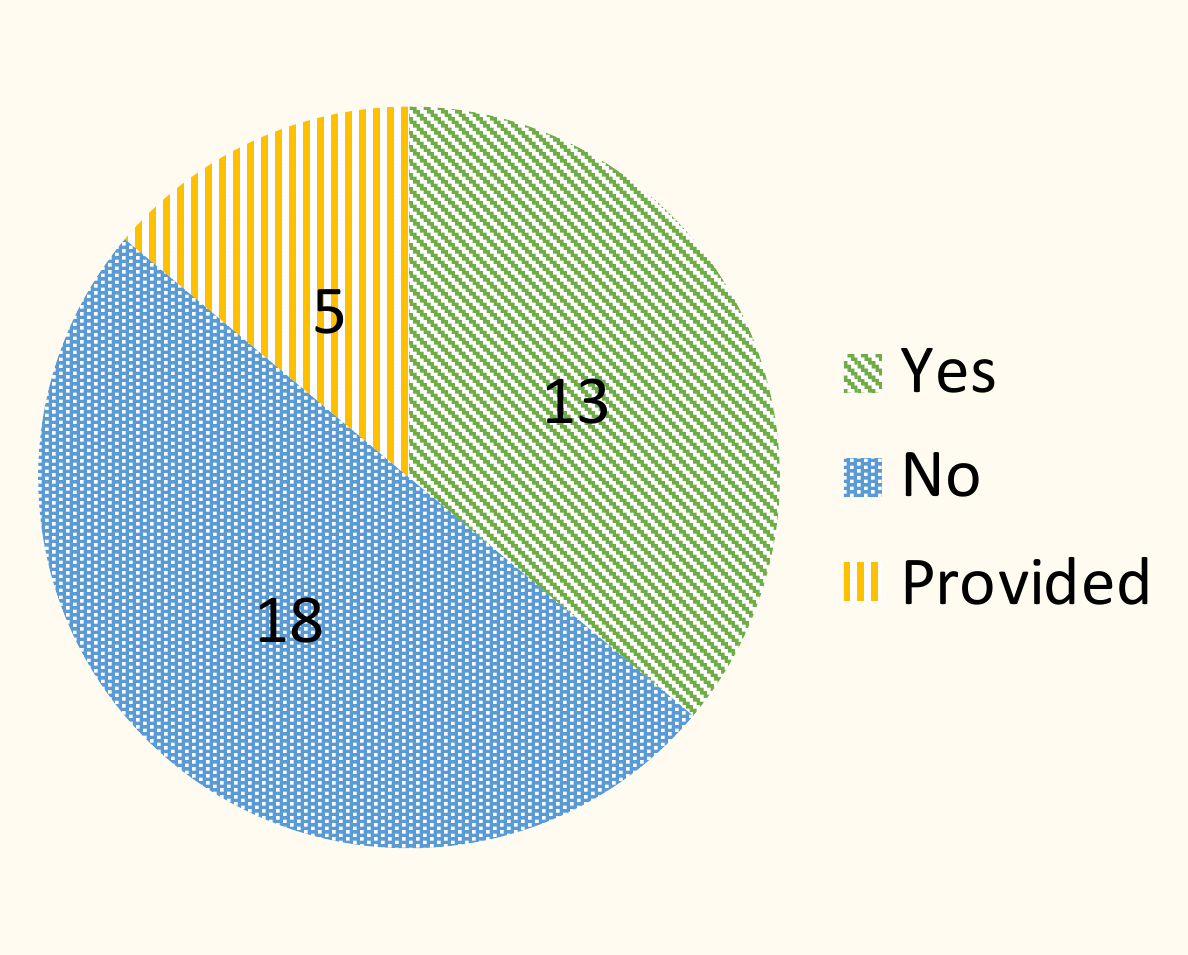}}
\caption{H1 Distribution}
\label{fig:h1}
\end{figure}

\begin{figure}[h]
\centerline{\includegraphics[width=.95\linewidth]{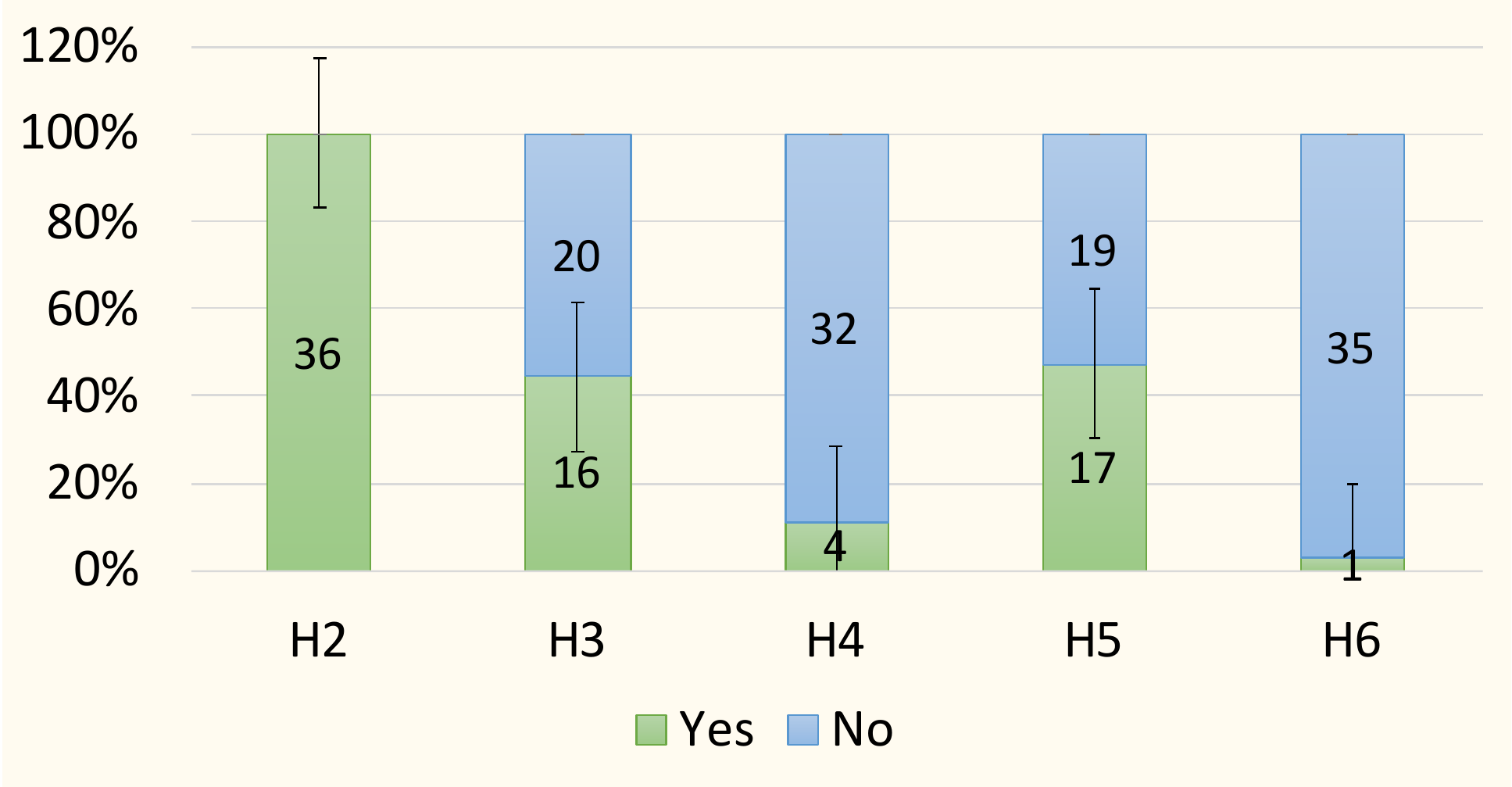}}
\caption{Satisfied Heuristics in Selected Websites}
\label{fig:obs}
\end{figure}

Now, we present an analysis of the findings of our study for each heuristic. Our findings for all heuristics have been summarised in Table \ref{tab:findings} where the symbol ``\CIRCLE'' has been used (for all heuristics except H1) to denote if a particular website satisfies a respective heuristic or the symbol ``\Circle'' has been used to denote the opposite (not satisfied) whereas \textit{PS} implies that the password is provided by a website. We have counted the number of heuristics satisfied by each investigated website and the distribution of satisfied heuristics are presented in Figure \ref{fig:h1} for H1 and Figure \ref{fig:obs} for other heuristics.

\newcolumntype{P}[1]{>{\centering\arraybackslash}p{#1}}
\setlength{\textfloatsep}{2pt}
\begin{table*}[h]
\caption{Detailed Information of 36 Websites}
\begin{center}
 \begin{tabular}{p{4.5cm}|p{4cm}|p{3.7cm}|P{0.6cm}|P{0.27cm}|P{0.27cm}|P{0.27cm}|P{0.27cm}|P{0.27cm}} 

 \hline
 \rowcolor[gray]{.9}\centering \textbf{Website} & \centering\textbf{URL} & \centering\textbf{Purpose} & \centering\textbf{H1} & \textbf{H2} & \textbf{H3}& \textbf{H4} & \textbf{H5}&\textbf{H6}\\ [0.5ex] 
 \hline\hline
 E challan & \url{http://echallan.gov.bd/login}  & Online challan service & \Circle & \CIRCLE & \CIRCLE & \Circle & \Circle & \Circle\\ 
 \hline
  
 \rowcolor[gray]{.96}Travel Agency Management System &\url{https://regtravelagency.gov.bd } & Agency registration service & PS & \CIRCLE & \Circle& \Circle& \CIRCLE & \Circle
\\
 \hline
 
  Police Clearance & \url{http://pcc.police.gov.bd/} & Police clearance service & PG-6 & \CIRCLE & \CIRCLE & \Circle & \Circle & \Circle  \\
 \hline
\rowcolor[gray]{.96}eRecruitment System & \url{https://erecruitment.bcc.gov.bd/exam/users/login}&  Online Recruitment service & PS & \CIRCLE & \CIRCLE & \Circle & \CIRCLE & \Circle \\
 \hline
 
 Bangladesh Investment Development Authority & \url{https://osspid.org/user/create} & Investor registration service & PS & \CIRCLE & \CIRCLE & \Circle & \CIRCLE & \Circle\\ 
 \hline
 \rowcolor[gray]{.96}Bangladesh Scouts & \url{http://service.scouts.gov.bd/login} & Member registration & \Circle & \CIRCLE & \CIRCLE & \Circle & \Circle & \Circle\\
 \hline
 
 eTin & \url{https://secure.incometax.gov.bd/TINHome} & eTin registration service & PG-4 & \CIRCLE & \CIRCLE & \CIRCLE & \CIRCLE & \Circle\\
 \hline
 \rowcolor[gray]{.96}Directorate General of Drug Administration &  \url{https://www.dgda.gov.bd/} & Drug regulatory authority service & \Circle & \CIRCLE & \Circle & \Circle & \CIRCLE & \Circle\\
 \hline
 eVeterinary & \url{http://evet.gov.bd/registration} & Veterinary information & \Circle & \CIRCLE & \Circle & \Circle & \Circle & \Circle\\
 \hline
 \rowcolor[gray]{.96}Dhaka Power Distribution Company Limited &  \url{https://dpdc.org.bd/career/} & Career portal service & PG-3 & \CIRCLE & \Circle & \Circle & \CIRCLE & \Circle\\
 \hline
 ePayment & \url{http://nbrepayment.gov.bd/} & Tax service & PG-1 & \CIRCLE & \CIRCLE & \CIRCLE & \Circle & \CIRCLE\\
 \hline
 \rowcolor[gray]{.96}NBR - learning & \url{http://nbrelearning.gov.bd/page/sign-up} & eLearning information & \Circle & \CIRCLE & \Circle & \Circle & \Circle & \Circle\\
 \hline
 EkSheba & \url{https://eksheba.gov.bd/} & Integrated service portal & PS & \CIRCLE & \Circle & \Circle & \CIRCLE & \Circle\\
 \hline
 \rowcolor[gray]{.96}National e-Government Procurement &  \url{https://www.eprocure.gov.bd/} & Govt. procurement service & PG-6 & \CIRCLE & \CIRCLE & \CIRCLE & \CIRCLE & \Circle\\
 \hline
 Biman Airlines & \url{https://www.biman-airlines.com/member/registration} & Online booking service & \Circle & \CIRCLE & \Circle & \Circle & \CIRCLE & \Circle\\
 \hline
 \rowcolor[gray]{.96}VAT  & \url{https://www.vat.gov.bd/} & Tax return service & PS & \CIRCLE & \Circle & \CIRCLE & \CIRCLE & \Circle\\
 \hline

 Bangladesh Telecommunication Regulatory commision & \url{https://naid.btrc.gov.bd/} & IMEI checking \& NOC information & \Circle & \CIRCLE & \Circle & \Circle & \CIRCLE & \Circle\\
 \hline
 \rowcolor[gray]{.96}Bangladesh Customs & \url{http://103.48.18.166/signup} & Auction service & PG-5
 & \CIRCLE & \Circle & \Circle & \Circle & \Circle\\
 \hline
 Trade License & \url{http://www.etradelicense.gov.bd/DefaultEng} & Trade license service & \Circle & \CIRCLE & \Circle & \Circle & \Circle & \Circle\\
 \hline
 \rowcolor[gray]{.96}Passport & \url{http://passport.gov.bd/Application-1.aspx} & Passport application service & \Circle & \CIRCLE & \Circle & \Circle & \Circle & \Circle\\
 \hline
 Teletalk & \url{http://www.teletalk.com.bd/} & Telecommunication service & PG-5 & \CIRCLE & \CIRCLE & \Circle & \Circle & \Circle\\
 \hline
 \rowcolor[gray]{.96}Rajuk & \url{http://cp.rajuk.gov.bd/user/signup} & Land/Construction permit & PG-2 & \CIRCLE & \CIRCLE & \Circle & \Circle & \Circle\\
 \hline
 eFire License & \url{http://efirelicense.gov.bd/apply-license-renew} & Fire license service & \Circle & \CIRCLE & \Circle & \Circle & \Circle & \Circle\\
 \hline
 \rowcolor[gray]{.96}Chief controller of import and export & \url{https://olm.ccie.gov.bd/login} & License approval \& renewal & \Circle & \CIRCLE & \CIRCLE & \Circle & \CIRCLE & \Circle\\
 \hline
 Mukhtopaath & \url{http://www.muktopaath.gov.bd/register} & e-Learning platform service & PG-4 & \CIRCLE & \Circle & \Circle & \Circle & \Circle\\
 \hline
 \rowcolor[gray]{.96}BCC-CA &  \url{https://www.bcc-ca.gov.bd/} & Certifying authority service  & PG-7 & \CIRCLE & \CIRCLE & \Circle & \CIRCLE & \Circle\\
 \hline
 Dokkhota Batayon & \url{http://skills.gov.bd/login} & Work-placement service & \Circle & \CIRCLE & \Circle & \Circle & \CIRCLE & \Circle\\
 \hline
 \rowcolor[gray]{.96}North west power generations company & \url{http://career.nwpgcl.gov.bd/} & Online job application service & \Circle & \CIRCLE & \Circle & \Circle & \Circle & \Circle\\
 \hline
 Skill connect &  \url{http://portal.bdskills.gov.bd} & Training portal & \Circle & \CIRCLE & \Circle & \Circle & \Circle & \Circle\\
 \hline
 \rowcolor[gray]{.96}Bangladesh Road Transport Authority &  \url{https://www.ipaybrta.cnsbd.com/index/carowner} & Vehicle fees and taxes payment service & PG-4 & \CIRCLE & \CIRCLE & \Circle & \CIRCLE & \Circle\\
 \hline
 Shikkhok batayon & \url{https://www.teachers.gov.bd/} & Blog & PG-2 & \CIRCLE & \Circle & \Circle & \CIRCLE & \Circle\\
 \hline
 
\rowcolor[gray]{.96} Agni Nirbapon & \url{http://www.noc.fireservicebd.info/auth/register} & Exemption certificate service & \Circle & \CIRCLE & \CIRCLE & \Circle & \Circle & \Circle\\
 \hline
 BAERA  &  \url{https://ells.baera.gov.bd/} & Nuclear control program authority info. & \Circle & \CIRCLE & \CIRCLE & \Circle & \CIRCLE & \Circle\\
 \hline
 \rowcolor[gray]{.96}Bangbandhu Sc. \& Tech. Fellowship &  \url{http://fellowship.bstft.gov.bd/} & Fellowship application service & PG-4 & \CIRCLE & \Circle & \Circle & \Circle & \Circle\\
 \hline
 Bostro Odhidoptor &  \url{http://e-application.dot.gov.bd/public/public/home} & e-Service for textile department & \Circle & \CIRCLE & \Circle & \Circle & \Circle & \Circle\\
 \hline
 \rowcolor[gray]{.96}NID & \url{https://services.nidw.gov.bd/nid-pub/citizen-home/} & NID registration, correction service  & \Circle & \CIRCLE & \CIRCLE & \Circle & \Circle & \Circle\\
 \hline
 
\end{tabular}
\label{tab:findings}
\end{center}
\vspace{-4mm}
\end{table*}

\setlength{\textfloatsep}{2pt}

\setlength{\textfloatsep}{2pt}

\noindent \textbf{H1:} From the selected websites, $13$ websites provide and enforce a password construction guideline (Figure \ref{fig:h1}). $18$ websites do not provide any guidelines and allow a password of any length and strength to be created. On the remaining $5$ websites, the password for a user is created by the system which is then supplied to the user. Three out of these five websites utilise an OTP (One Time Password) sent over another channel (e.g. email, mobile phone and so on) during the registration as well login phases instead of using a fixed password. This is an alternative mode of authentication, however, the user might face severe consequences when the user loses control of the registered phone number or email address. Among the $13$ websites providing guidelines, they have adopted a wide variety of guidelines. We have grouped these guidelines in seven different categories as presented below, where \textit{PG} stands for \textit{Password Guideline}. Out of these guidelines, PG-6 is supposedly more secure if no dictionary word is used. PG-1 can be the least secure one as it allows a password of only four characters without any mixture of letters, numbers and special characters. Even PG-7 can be vulnerable since it allows any password length.
\begin{itemize}
    \item \textbf{PG-1}: At least 4 characters
    \item \textbf{PG-2}: At least 6 characters
    \item \textbf{PG-3}: At least 6 characters with a mixture of letters and numbers
    \item \textbf{PG-4}: At least 8 characters
    \item \textbf{PG-5}: At least 8 characters with a mixture of letters and numbers
    \item \textbf{PG-6}: At least 8 characters with a mixture of letters, numbers and special characters
    \item \textbf{PG-7}: A mixture of letters, numbers and special characters with no specified length
\end{itemize}

\begin{figure}[h]
\thispagestyle{empty}
\centerline{\includegraphics[width=0.96\linewidth]{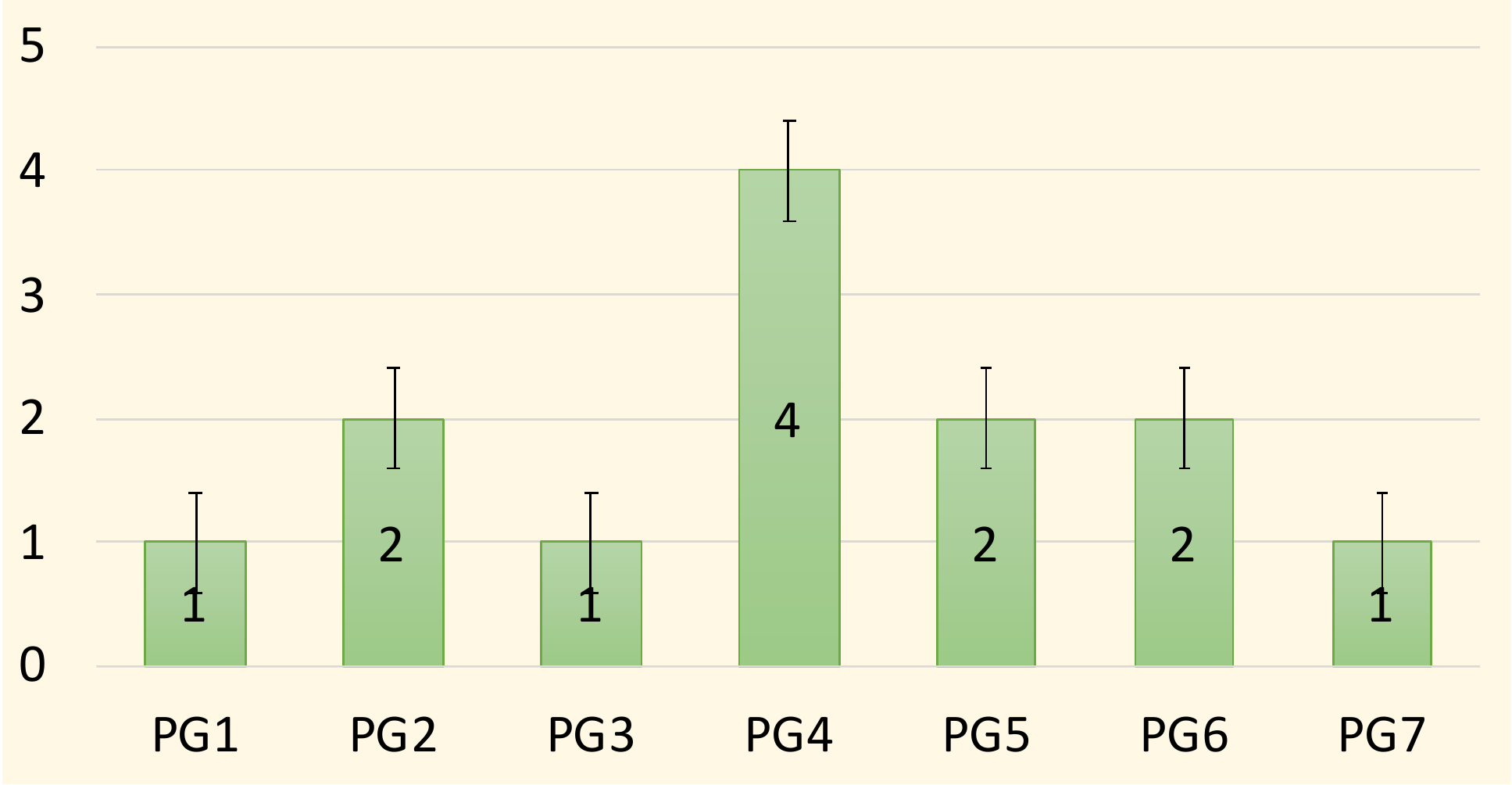}}
\caption{Distribution of Password Guidelines}
\label{fig:pg}
\end{figure}

The corresponding guideline has been used in the respective field for H1 column in Table \ref{tab:findings}. The distribution of these guidelines among the $13$ websites are presented in Figure \ref{fig:pg}. As per the figure, the mostly adopted guidelines is PG-4 requiring a password length of 8 characters adopted by $4$ websites.

\vspace{2mm}
\noindent \textbf{H2:} Password recovery factor being a basic security mechanism, has been found in all $36$ websites (Figure \ref{fig:obs}). However, the interesting observation is the recovery channel utilised by different websites. For example, $24$ websites (more than $66\%$) have adopted emails as their password recovery channel into which the password reset links are sent. Such links are valid for a specific amount of time and passwords must be reset within that validity period. Mobile phones have been adopted as the password recovery channel by $7$ website where an OTP is sent to reset the password. The remaining $5$ websites use both of these channels for password recovery.

\vspace{2mm}
\thispagestyle{empty}
\noindent \textbf{H3:} As reported in Table \ref{tab:findings} and Figure \ref{fig:obs}, CAPTCHA input is mandatory during the registration process on $16$ websites while the rest of the $20$ websites do not employ any CAPTCHA. As explained earlier, CAPTCHA might guard against the creation of falsified accounts via automated computers programs. Therefore, it is highly recommended to employ CAPTCHA and the websites which do not utilise CAPTCHA are susceptible against such attacks. 

\vspace{2mm}
\noindent \textbf{H4:} As per our investigation only $4$ websites (Figure \ref{fig:obs}) have asked security questions during the registration phase. Examples of some of the questions asked are:
\begin{itemize}
\item What is the name of your favourite book?
\item What is the name of your first school?
\item What is your mother’s maiden name?
\item What was your dream job as a child?
\item Who was your childhood hero?
\end{itemize}

Other $32$ websites has no security question, thus, an attacker can falsely and easily initiate an account recovery process. 

\vspace{2mm}
\noindent \textbf{H5:} It has been a surprise for us to observe that only $17$ websites (Figure \ref{fig:obs}) utilise an HTTPS channel, indicating that they are aware of the importance of a secure protocol. Unfortunately, the other $19$ websites rely on the insecure HTTP channel and hence, they remain extremely insecure, and every information, including passwords, transmitted to these websites might be visible to attackers during the transmission.

\vspace{2mm}
\noindent \textbf{H6:} Another surprising finding is that only one website (Figure \ref{fig:obs}) provides a password strength meter during the registration phase.  Such a visualisation feature helps any user to create a strong and more secure password. That is why it is crucial that any website requiring a password entry should implement a password strength meter. 

\subsection{Recommendations}
\label{sec:result:subsec:reco}
In this section, we present a number of recommendations to improve the security, with respect to passwords, of Bangladeshi government websites. 

\vspace{2mm}
\noindent \textbf{Recommendation-1:} It is imperative that every governmental website adopts and enforces a secure password construction guideline. This will ensure that users do not opt in for easy-to-guess insecure passwords. However, as we have found, there are a wide-range of password guidelines adopted by different governmental websites. To ensure a streamlined approach, we recommend to adopt (and show) a unified password construction guideline by (on) all governmental websites during the registration procedure. In addition to this, a password strength meter should be utilised to aid users selecting a secure, strong and compliant password.

\vspace{2mm}
\noindent \textbf{Recommendation-2:} Every governmental website must employ secure CAPTCHA mechanisms during the registration as well as password recovery phases. There are a wide-range of CAPTCHAs available varying in their types, security and usability. A unified policy for CAPTCHA should be devised and adopted for all governmental websites.

\vspace{2mm}
\noindent \textbf{Recommendation-3:} Each governmental website that deals with user registrations including password creation, storage and management must adopt an HTTPS-only policy to secure network traffic to and from their website.

\vspace{2mm}
\noindent \textbf{Recommendation-4:} The secure storage of passwords is also very crucial. In an ideal scenario, every single password should never be transmitted via the network to the server. Instead, they should be hashed at the user side before their transmission (resp. storage) to (in) the server. This reduces the risk of password leakage in case a breach occurs at the respective server. Alternatively, it must be ensured that passwords are transmitted through an HTTPS channel and stored in encrypted form in the backend database. 

\vspace{2mm}
\noindent \textbf{Recommendation-5:} Needing to create many accounts across different websites, a user will ultimately face the ill consequence of password fatigue \cite{das2014tangled}, a pressing feeling to remember and manage passwords for so many websites. This forces a user to reuse one or two secure passwords for different websites, thereby increasing the chance of attacks \cite{ives2004domino}. An effective technique to offset password fatigue in the governmental setting is to adopt the notion of an \textit{Identity Federation}. An identity federation is a trusted service model in which different service providers (known as \textit{SPs}) are tied together under a formal agreement with a single identity provider (\textit{IdP}) \cite{chadwick2009federated}. The IdP handles all identity activities such as identity creation, storage and management. Users need to register with that single IdP and they can access services from federated SPs using a single credential, thus significantly reducing the probability of password fatigue. Towards this aim, a proposal for a country-wide identity federation for Bangladesh was put forward by the authors in \cite{sadek2012identity}. Sadly, it was difficult to adopt this approach previously as there was no central identity database in Bangladesh. However, with the current National ID database and its associated service such as Porichoy (\texttt{https://porichoy.gov.bd/}), Bangladesh currently has the infrastructure to deploy a nation-wide identify federation. We highly recommend to adopt this approach.

\vspace{2mm}
\section{Conclusion}
\label{sec:conclusion}
In this paper, we have presented a study of password security among $36$ different Bangladeshi government websites against six carefully-chosen password security heuristics. The main motivation is to understand the password security status of the chosen websites which in a way acts as the reflection of the password security awareness and preparedness among the corresponding government agencies. Our findings suggest that many websites have not adopted proper security measures. Many of them do not utilise any password construction guideline which can act as a barrier for users to select a strong password. Even the security guidelines employed by a few websites are not so secure. Similarly, many websites still do not leverage the HTTPS capability to securely transmit information, indicating a lack of knowledge as well as preparedness with respect to security. We hope that our study sheds some lights on some important password security factors and the respective authority takes a note of their lacking and act accordingly. However, it must be pointed out that passwords are just a single aspect of the overall security of any website. A rigorous evaluation of the remaining aspects must be carried out to fully understand the comprehensive security status and preparedness of different Bangladeshi government websites which we aim to do in future. 

\bibliographystyle{IEEEtran}
\bibliography{main}

\thispagestyle{empty}
\end{document}